%% file: main.tex
\documentclass[sigconf,natbib=true,screen=true,anonymous=false,review=false]{acmart}

\input{preamble/authors}
\input{preamble/definitions}

\input{preamble/meta}

\begin{document}
\title{CLAX: Fast and Flexible Neural Click Models in JAX}



\input{sections/0-abstract}

\maketitle
\pagestyle{plain} 
\renewcommand\footnotetextcopyrightpermission[1]{} 

\input{sections/1-introduction.tex}
\input{sections/2-related-work.tex}
\input{sections/3-em.tex}
\input{sections/4-clax.tex}
\input{sections/5-numerical-stability.tex}
\input{sections/6-experiments.tex}
\input{sections/7-results.tex}
\input{sections/8-conclusion.tex}
\input{sections/9-acknowledgements.tex}

\bibliographystyle{ACM-Reference-Format}
\bibliography{main}

\input{sections/10-appendix.tex}

\end{document}

%% file: preamble/authors.tex
\author{Philipp Hager}
\orcid{0000-0001-5696-9732}
\affiliation{%
    \institution{University of Amsterdam}
    \city{Amsterdam}
    \country{The Netherlands}
}
\email{p.k.hager@uva.nl}

\author{Onno Zoeter}
\orcid{0000-0003-1704-706X}
\affiliation{%
    \institution{Booking.com}
    \city{Amsterdam}
    \country{The Netherlands}
}
\email{onno.zoeter@booking.com}

\author{Maarten de Rijke}
\orcid{0000-0002-1086-0202}
\affiliation{%
    \institution{University of Amsterdam}
    \city{Amsterdam}
    \country{The Netherlands}
}
\email{m.derijke@uva.nl}

%% file: preamble/definitions.tex
\usepackage[inline]{enumitem}
\usepackage{mathtools}
\usepackage{derivative}

\usepackage{multirow}
\usepackage{acronym}
\usepackage{stfloats}
\usepackage[skip=0pt]{caption}
\usepackage{etoolbox}
\usepackage{dsfont}
\usepackage{listings}
\usepackage{xcolor}

\definecolor{keywordcolor}{RGB}{31,119,180}

\makeatletter 
\newcommand\mynobreakpar{\par\nobreak\@afterheading} 
\makeatother



%% file: sections/0-abstract.tex
\begin{abstract}
CLAX is a JAX-based library that implements classic click models using modern gradient-based optimization. While neural click models have emerged over the past decade, complex click models based on probabilistic graphical models (PGMs) have not systematically adopted gradient-based optimization, preventing practitioners from leveraging modern deep learning frameworks while preserving the interpretability of classic models. CLAX addresses this gap by replacing EM-based optimization with direct gradient-based optimization in a numerically stable manner. The framework's modular design enables the integration of any component, from embeddings and deep networks to custom modules, into classic click models for end-to-end optimization. We demonstrate CLAX's efficiency by running experiments on the full Baidu-ULTR dataset~\cite{Zou2022Baidu} comprising over a billion user sessions in $\approx$ 2 hours on a single GPU, orders of magnitude faster than traditional EM approaches. CLAX implements ten classic click models, serving both industry practitioners seeking to understand user behavior and improve ranking performance at scale and researchers developing new click models. CLAX is available at: \url{https://github.com/philipphager/clax}
\end{abstract}

%% file: sections/1-introduction.tex
\section{Introduction}
\label{sec:introduction}

Click models are an integral part of web search and recommender systems, serving as fundamental tools for understanding and predicting user interactions~\cite{Chuklin2015ClickModels}. Click models are used to determine user behavior on search engine result pages~\cite{Richardson2007PBM,Craswell2008PBM,Dupret2008UBM,Chapelle2009DBN}, predict clicks in advertising~\cite{Chen2012PosNormAds,McMahan2013AdsTrenches,Zhu2010AdClicks}, estimate click biases for counterfactual learning-to-rank methods~\cite{Wang2018RegressionEM,Vardasbi2020Affine}, simulate users in reinforcement learning~\cite{Deffayet2024Sardine, Huang2020RLSimulation}, evaluate new ranking systems~\cite{Chuklin2013ClickMetrics}, and are directly used as ranking models~\cite{Yan2022TwoTowers}.

\begin{figure}
\begin{lstlisting}[language=Python, label={lst:ubm-example}, caption={A minimal example of training a UBM in CLAX.}]
from clax import Trainer, UserBrowsingModel
from flax import nnx
from optax import adamw

model = UserBrowsingModel(
    query_doc_pairs=100_000_000,
    positions=10,
    rngs=nnx.Rngs(42),
)
trainer = Trainer(
    optimizer=adamw(0.003),
    epochs=50,
)
train_df = trainer.train(model, train_loader, val_loader)
test_df = trainer.test(model, test_loader)
\end{lstlisting}
\vspace*{-2em}
\end{figure}

Early click models can be represented as probabilistic graphical models (PGMs)~\cite{Chuklin2015ClickModels} that explicitly model variables such as document attractiveness~\cite{Richardson2007PBM,Craswell2008PBM}, user satisfaction~\cite{Chapelle2009DBN,Guo2009CCM}, and rank examination~\cite{Richardson2007PBM,Dupret2008UBM} that might influence user behavior. Since most of these variables are latent, popular models including the position-based model (PBM)~\cite{Richardson2007PBM}, user browsing model (UBM)~\cite{Dupret2008UBM}, and dynamic Bayesian network (DBN) of \cite{Chapelle2009DBN} are optimized using expectation maximization (EM). EM iteratively optimizes probabilistic models with missing variables by alternately imputing missing values and updating parameters, and is widely employed in click modeling libraries like PyClick\footnote{\url{https://github.com/markovi/PyClick}} or ParClick~\cite{Khandel2022ParClick}. The iterative nature of EM, however, scales poorly with dataset size, motivating specialized algorithms~\cite{Khandel2022ParClick,Thijssen2023MassiveClicks} and online optimization procedures~\cite{Markov2017OnlineEM}.

Over the last decade, neural click models have emerged along two directions. The first uses neural architectures such as recurrent neural networks~\cite{Borisov2016NCM,Chen2020CACM}, attention~\cite{Zhuang2021XPA}, or graph neural networks~\cite{Lin2021GraphCM} to improve click prediction and ranking performance, potentially sacrificing the interpretability that made traditional PGMs valuable for understanding users. The second branch parameterizes simple PGMs with deep neural networks~\cite{Yan2022TwoTowers,Deffayet2023Robustness}. A prominent example is the two-tower model, a neural parameterization of the PBM~\cite{Richardson2007PBM}, popular for position bias correction in industry~\cite{Guo2019PAL,Haldar2020Airbnb,Yan2022TwoTowers,Zhao2019AdditiveTowers}.

However, a significant gap remains: \emph{the paradigm shift to gradient-based optimization has not been systematically applied to more complex PGM-based click models~\cite{Chuklin2015ClickModels}, preventing practitioners from leveraging modern deep learning frameworks while preserving the interpretability and theoretical foundations of classic models}. Given that simple two-tower models already demonstrate strong empirical performance~\cite{Guo2019PAL,Haldar2020Airbnb,Yan2022TwoTowers}, parameterizing more sophisticated models like the DBN~\cite{Chapelle2009DBN} and UBM~\cite{Dupret2008UBM} may yield significant improvements.

Modern frameworks such as JAX~\cite{Bradbury2018Jax} are compelling for addressing this gap. JAX's functional programming and automatic differentiation are well-suited for complex probabilistic models, while just-in-time (JIT) compilation and vectorization enable efficient computation on large datasets. And high-level deep learning libraries, such as Flax NNX~\cite{Heek2020Flax}, make JAX more accessible.

We introduce CLAX: A neural click modeling library bridging traditional PGMs with modern gradient-based optimization in JAX. CLAX replaces EM with direct optimization of marginal log-likelihoods using gradient descent in a numerically stable way.

CLAX is modular and allows broad customization. By default, CLAX uses embeddings, e.g., to represent the attractiveness of individual query-document pairs, making it equivalent to classic click modeling libraries. Listing~\ref{lst:ubm-example} shows an example of training a UBM with CLAX. Building on embeddings, CLAX uses methods from the deep recommender systems literature to compress large embedding tables~\cite{Weinberger2009HashingTrick,Shi2020CompositionalEmbeddings}, enabling scaling to billions of query-document pairs on a single GPU. Importantly, CLAX enables flexible customizations beyond embeddings, including linear models, deep networks, deep-cross networks, and custom FLAX models. Any module matching the output shapes expected by CLAX can be plugged in for end-to-end optimization. This modularity even allows meta-models, which we demonstrate by building on ideas by \citet{Yan2022TwoTowers} and training a mixture model over multiple click models for datasets where users exhibit different behaviors across sessions. To summarize, our contributions are fourfold:
\begin{itemize}[nosep, leftmargin=*]
    \item We demonstrate that direct gradient-based optimization can replace EM for training traditional PGM-based click models and achieves comparable empirical performance.
    \item We introduce CLAX, the first JAX-based click modeling library with a highly modular design allowing any component to be plugged into classic PGM structures.
    \item We showcase the computational efficiency of CLAX by training and evaluating on over 1B user sessions of the Baidu-ULTR dataset on a single GPU.
    \item We show that neural parameterizations of sophisticated PGMs can surpass the ranking performance of widely-used two-tower models, providing a powerful new tool for practitioners.
\end{itemize}
CLAX is meant as a library for practitioners and researchers alike. Industry practitioners might extend two-tower models and use cascade-like models. Researchers benefit from the demonstration of how the marginal log-likelihood of many PGMs can be optimized directly using SGD and autograd frameworks, simplifying the creation of new models. The core paradigm of CLAX is not tied to JAX. All models can be implemented in PyTorch or TensorFlow. We chose JAX for its computational speed and growing ecosystem.

Below, we first cover related work on click modeling, implementations in the field, and the JAX ecosystem. Section~\ref{sec:em} compares gradient-based with EM-based optimization for click models. Section~\ref{sec:clax} provides an overview of the CLAX library and Section~\ref{sec:numerical-stability} discusses the basics of its numerically stable implementation. Lastly, we evaluate CLAX empirically in Sections~\ref{sec:experiments} and \ref{sec:results}.

%% file: sections/2-related-work.tex
\section{Related Work}
\label{sec:related-work}

\subsection{Click models}
Click models emerged as probabilistic graphical models (PGMs) to predict user clicks on search engine result pages~\cite{Chuklin2015ClickModels}. Most click models are extensions of two models. The position-based model (PBM)~\cite{Richardson2007PBM} assumes that users click after examining the position of a document and finding the document attractive. Notably, observing and clicking on a document under the PBM is independent of all other documents in the same ranking. The cascade model~\cite{Craswell2008PBM} is the second foundational model, assuming that users examine items sequentially from top to bottom, click on the first relevant item, and then stop browsing. Note that the cascade model can only explain a single click per result page. 

The dependent click model (DCM)~\cite{Guo2009DCM} extends the cascade model to account for multiple clicks by assuming a rank-dependent continuation probability, whereby users may continue browsing after clicking on a document. The click chain model (CCM)~\cite{Guo2009CCM} extends this idea by assuming users continue after a click based on the attractiveness of the current document. The dynamic Bayesian network (DBN)~\cite{Chapelle2009DBN} further extends this idea by separating document attraction (before click) and satisfaction (which is revealed after clicking). A popular extension, the SDBN assumes that users always continue browsing when they are not satisfied with the current item, whereas the DBN learns a separate continuation parameter.
The user browsing model (UBM)~\cite{Dupret2008UBM} extends the PBM by assuming that examining a document not only depends on the current position but also on the position of the last clicked document, thereby relaxing the independence assumption of the PBM. 

CLAX implements seven foundational PGM-based click models and three CTR-based baselines, as covered by \citet{Chuklin2015ClickModels}, using gradient-based optimization. We provide the complete derivation of each model and its marginal log-likelihood in Appendix~\ref{appendix:models}.

\vspace{-0.5em}
\subsection{Neural click models}
In recent years, click models have incorporated neural networks in two ways. First, new click models based on neural architectures have emerged. Models like the NCM \cite{Borisov2016NCM} or the CACM \cite{Chen2020CACM} predict clicks using recurrent neural networks. The XPA model by \citet{Zhuang2021XPA} uses attention to capture interactions between documents in a grid layout. And the GraphCM~\cite{Lin2021GraphCM} leverages graph neural networks to model user behavior across sessions. These methods demonstrate strong empirical performance in click prediction at the risk of being less interpretable than PGM-based click models. While these models can be implemented with Flax NNX and CLAX, our focus in this work is on gradient-based optimization of classic PGM-based models.

Second, neural implementations of classic click models have emerged. Neural networks based on the PBM are known as two-tower models and have become prevalent in industry ranking settings~\cite{Guo2019PAL,Haldar2020Airbnb,Yan2022TwoTowers}, allowing the use of document features and custom network architectures to predict examination and attractiveness. CLAX provides the logical next step by enabling easy parameterization of more advanced PGM-based click models.

\vspace{-0.5em}
\subsection{Frameworks for click modeling}
The standard library for click modeling is PyClick,\footnote{\url{https://github.com/markovi/PyClick}} published by \citet{Chuklin2015ClickModels}, which optimizes click models using maximum likelihood estimation and EM. While their iterative EM optimization can be considerably accelerated with the PyPy\footnote{\url{https://www.pypy.org/}} interpreter, it remains too slow even for medium-scale datasets (see Section~\ref{sec:results:yandex}). To address these performance limitations, specialized libraries have emerged. ParClick~\cite{Khandel2022ParClick}\footnote{\url{https://github.com/uva-sne/ParClick}} is a C++ library that parallelizes EM across multiple CPU cores on a single machine. MassiveClicks~\cite{Thijssen2023MassiveClicks}\footnote{\url{https://github.com/skip-th/MassiveClicks}} extends ParClick to support multi-node and GPU-based training, achieving substantial improvements in training time. However, these specialized libraries focus primarily on EM-style optimization, making them difficult to extend with custom neural modules or use additional features.

Rather than competing with specialized libraries on computational speed, CLAX takes a different approach. We implement all PyClick models, replacing EM with gradient-based optimization using a general-purpose deep learning framework. This enables the end-to-end optimization of fully customizable models, while benefiting from JAX's speed through just-in-time compilation and GPU support~\cite{Bradbury2018Jax,Heek2020Flax}. Our JAX implementation demonstrates a paradigm that can be translated to other deep learning frameworks.

Lastly, we acknowledge that we are not the first to apply gradient-based optimization to PGM-based click models beyond PBM. \citet{Deffayet2023Robustness} implement the PBM, UBM, and DBN using gradient-based optimization in PyTorch while investigating click model robustness to distributional shift. However, since their work focused on robustness analysis rather than providing a general-purpose library, CLAX offers a more extensible API and improved numerical stability through computation in log-probability space.

\vspace{-0.5em}
\subsection{The JAX ecosystem}
JAX is a Python library for numerical computation built around composable function transformations that enable automatic differentiation, JIT-compilation to GPU/TPUs, and support for vectorized and distributed computing~\cite{Bradbury2018Jax}. Unlike monolithic frameworks such as PyTorch or TensorFlow, JAX adopts a modular ecosystem approach where functionality is distributed across specialized libraries~\cite{DeepMind2020JAXEcosystem}. The base JAX library provides a NumPy-compatible interface and fundamental transformations, while additional libraries offer domain-specific functionality: Flax NNX~\cite{Heek2020Flax} and Haiku~\cite{Hennigan2020Haiku} are neural network libraries, Optax supplies optimizers, Chex offers testing utilities, and Distrax provides probability distributions. This modular design enables research communities to develop specialized tools for their respective field~\cite{Kidger2021Diffrax,Rader2023Lineax,Cabezas2024BlackJax}. Within information retrieval, Rax~\cite{Jagerman2022Rax} is the primary library for learning-to-rank loss functions and evaluation metrics. CLAX is the first click modeling framework in the ecosystem, integrating with Optax for optimization, Flax NNX for deep learning, and Rax for ranking metrics.

%% file: sections/3-em.tex
\section{Comparing Expectation Maximization and Gradient-based Optimization}
\label{sec:em}

Next, we compare the expectation maximization algorithm~\cite{Dempster1977EM} and gradient ascent\footnote{In this section, we frame the problem as likelihood maximization, although in practice we minimize the negative log-likelihood using gradient descent.} for inferring parameters in click models, using the example of the position-based model (PBM)~\cite{Richardson2007PBM,Craswell2008PBM}. We focus our analysis on a single query with documents $d \in D$ displayed at positions $k \in \{1, \dots, K\}$ to simplify our notation. The PBM assumes that a user clicks on a document $d$ if they examine its position $k$ and find it attractive.\footnote{More commonly in connection with the PBM is the term ``relevant.'' But as the click models in this work differentiate between users being attracted to a document snippet and being satisfied with a document after clicking, we follow \citet{Chuklin2015ClickModels} and use the more precise terminology of ``attractiveness.''}. We define binary random variables for click $C$, examination $E$, and attractiveness $A$, with realizations $c, e$ and $a$. The click probability of the PBM is:
\begin{equation}
    P(C = 1 \mid d, k) = P(E = 1 \mid k) \cdot P(A = 1 \mid d) = \theta_k \gamma_d,
\end{equation}
where $\theta_k$ is the probability of examining rank $k$ and $\gamma_d$ is the attractiveness probability of document $d$. Like many click models, the PBM contains latent variables. We only observe clicks, not whether a user examined a document or found it attractive. This means we cannot directly optimize the complete-data log-likelihood. Instead, we must find parameters that maximize the log-likelihood of the data we can actually observe, the marginal log-likelihood:
\begin{equation}
    \mathcal{L}(\theta, \gamma; \mathcal{D}) = \sum_{\mathclap{(d, k, c) \in \mathcal{D}}} \left[ c \log(\theta_k \gamma_d) + (1 - c) \log(1 - \theta_k \gamma_d) \right].
    \label{eq:marginal_likelihood}
\end{equation}

\subsection{Expectation-maximization (EM) algorithm}
A prominent method for optimizing likelihoods with latent variables is the EM algorithm~\cite{Dempster1977EM}. Starting from an initial guess for our model parameters, EM cycles between two steps until convergence. In the expectation step, we use the current model parameters and observed data to compute the posterior expectations of our latent variables. This step fills in the missing observations in our dataset. In the maximization step, we use these expected values to find new parameters that maximize the log-likelihood of the complete data. We use the newly obtained parameters to refine our posteriors in the next E-step, which leads us to find better fitting parameters in the next M-step, and so on. In the following, we showcase this principle for the PBM.

\subsubsection*{E-step:} Given parameters of the PBM $(\theta^{(t)}, \gamma^{(t)})$ at iteration step~$t$ and our observed clicks, we compute the posterior expectation of each latent variable for each observation $(d,k,c) \in \mathcal{D}$. For the binary variables of examination $E$ and attractiveness $A$, this is equivalent to their posterior probability given the observed click $c$:
\begin{align}
    \hat{e}_{d,k,c} &= \mathbb{E}[E \mid c, d, k; \theta^{(t)}, \gamma^{(t)}] \nonumber\\
    &= c \cdot P(E=1 \mid C=1) + (1-c) \cdot P(E=1 \mid C=0) \nonumber\\
    &= c \cdot 1 + (1-c) \cdot \frac{P(C=0\mid E=1, d, k) P(E=1 \mid k)}{P(C=0 \mid d,k)} \nonumber\\
    &= c + (1-c) \cdot \frac{(1 - \gamma_d^{(t)}) \theta_k^{(t)}}{1 - \theta_k^{(t)} \gamma_d^{(t)}},\\
    \hat{a}_{d,k,c} &= \mathbb{E}[A \mid c, d, k; \theta^{(t)}, \gamma^{(t)}] \nonumber\\
    &= c + (1-c) \cdot \frac{(1 - \theta_k^{(t)}) \gamma_d^{(t)}}{1 - \theta_k^{(t)} \gamma_d^{(t)}}.
\end{align}
\subsubsection*{M-step:} In the maximization step, we use the posterior expectations $(\hat{e}_{d,k,c}, \hat{a}_{d,k,c})$ computed for each observation in the E-step at time $t$ to find new model parameters by maximizing the expected complete-data log-likelihood ($\mathcal{Q}$-function):
\begin{equation}
    \begin{split}
    &\mathcal{Q}\left( \theta^{(t + 1)}, \gamma^{(t + 1)} \mid \theta^{(t)}, \gamma^{(t)} \right) =\\
    &\sum_{\mathclap{(d, k, c) \in \mathcal{D}}} \left[ \hat{e}_{d,k,c} \log(\theta^{(t + 1)}_k) + (1 - \hat{e}_{d,k,c}) \log(1 - \theta^{(t + 1)}_k) \right] +\\
    &\sum_{\mathclap{(d, k, c) \in \mathcal{D}}} \left[ \hat{a}_{d,k,c} \log(\gamma^{(t + 1)}_d) + (1 - \hat{a}_{d,k,c}) \log(1 - \gamma^{(t + 1)}_d) \right].
    \end{split}
\end{equation}
Note that the $\mathcal{Q}$-function is commonly much simpler than our marginal log-likelihood. In the case of the PBM, it allows us to decouple the estimation of $\theta$ and $\gamma$. By taking the derivative of $\mathcal{Q}$ with respect to each parameter, setting it to zero, and solving, we obtain closed-form update rules for the PBM:
\begin{equation}
    \begin{split}
        \theta_k^{(t+1)} &= \frac{\sum_{(d,k',c) \in \mathcal{D}, k'=k} \hat{e}_{d,k,c} }{\sum_{(d,k',c) \in \mathcal{D}, k'=k} 1},\\
        \gamma_d^{(t+1)} &= \frac{\sum_{(d',k,c) \in \mathcal{D}, d'=d} \hat{a}_{d,k,c} }{\sum_{(d',k,c) \in \mathcal{D}, d'=d} 1},\\
    \end{split}
\end{equation}
which divides the sum of all expected posterior examination values by the number of documents at a given position, and the sum of all expected attractiveness values for a document by the number of impressions of that document.

\vspace{-0.5em}
\subsection{Gradient-based optimization}
An alternative to EM is to optimize the marginal log-likelihood $\mathcal{L}$ in Eq.~\ref{eq:marginal_likelihood} directly using gradient-based methods. This involves computing the partial derivative of $\mathcal{L}$ with respect to each parameter and taking a step in the direction of the gradient~\cite{Ruder2017GradientDescent}:
\begin{equation}
    \frac{\partial \mathcal{L}}{\partial \gamma_d} = \sum_{(d',k,c) \in \mathcal{D}, d'=d} \left( \frac{c}{\gamma_{d}} - \frac{(1 - c) \theta_k}{1 - \theta_k \gamma_{d}} \right),
\end{equation}
\begin{equation}
    \frac{\partial \mathcal{L}}{\partial \theta_k} = \sum_{(d,k',c) \in \mathcal{D}, k'=k} \left( \frac{c}{\theta_{k}} - \frac{(1 - c) \gamma_d}{1 - \theta_{k} \gamma_d} \right).
\end{equation}
We update model parameters iteratively using learning rate $\eta$:
\begin{equation}
    \theta_k^{(t+1)} = \theta_k^{(t)} + \eta \frac{\partial \mathcal{L}}{\partial \theta_k}
    \quad \text{and} \quad
    \gamma_d^{(t+1)} = \gamma_d^{(t)} + \eta \frac{\partial \mathcal{L}}{\partial \gamma_d}.
    \label{eq:gd_updates}
\end{equation}

\subsection{Comparison and discussion}
Both EM and gradient-based optimization are valid methods for finding maximum likelihood estimates when dealing with marginal log-likelihoods~\cite[Chapter 19]{Koller2009PGM}. While both can become trapped in local optima, they differ in their optimization characteristics. EM guarantees monotonic improvement in the marginal log-likelihood at each iteration, and it circumvents the complexity of directly optimizing the marginal likelihood by optimizing a simpler auxiliary function~\cite{Dempster1977EM,Wu1983EMConvergence,Salakhutdinov2003EM}. However, EM can be slow to converge in settings with high missing information~\cite{Wu1983EMConvergence,Dempster1977EM}, and classical implementations require full dataset passes, a limitation that motivated online and stochastic variants~\cite{Neal1998IncrementalEM,Thiesson2001AcceleratingEM,Cappe2011OnlineEM}.

Gradient-based methods offer different trade-offs, using general-purpose optimization advances (e.g., adaptive learning rates~\cite{Duchi2011AdaGrad,Kingma2014Adam,Loshchilov2019AdamW}, momentum~\cite{Qian1999Momentum}) that can lead to fast convergence, without however, the guarantee of monotonic improvements that EM provides. Their key practical advantage lies in mini-batch processing, which scales well to large datasets and modern parallel hardware. The tractability of marginal log-likelihoods of click models makes gradient optimization particularly attractive, as automatic differentiation eliminates the need for manual E and M step derivations while potentially offering computational efficiency gains.

Lastly, the relationship between EM and gradient methods runs deeper than their shared objective. A fundamental property links the two approaches: the gradient of the expected complete-data log-likelihood ($\mathcal{Q}$-function), evaluated at the current parameter estimates, is equal to the gradient of the marginal log-likelihood~\cite{Salakhutdinov2003EM}:
\begin{align}
    \nabla \mathcal{Q}(\theta, \gamma \mid \theta^{(t)}, \gamma^{(t)})\Big|_{\theta=\theta^{(t)}, \gamma=\gamma^{(t)}} 
    = \nabla \mathcal{L}(\theta, \gamma) \Big|_{\theta=\theta^{(t)}, \gamma=\gamma^{(t)}}.
\end{align}
For the PBM, we can verify this equality explicitly. The gradient of the auxiliary function with respect to $\gamma_d$, evaluated at $\theta_k = \theta^{(t)}_k$ and $\gamma_d = \gamma^{(t)}_d$, simplifies to:
\begin{equation}
    \frac{\partial \mathcal{Q}\left( \theta_k, \gamma_d \mid \theta_k, \gamma_d \right)}{\partial \gamma_d} = \hspace{-12pt} \sum_{(d',k,c) \in \mathcal{D}, d'=d} \left( \frac{c}{\gamma_{d}} - \frac{(1 - c) \theta_k}{1 - \theta_k \gamma_{d}} \right) = \frac{\partial \mathcal{L}}{\partial \gamma_d}.
\end{equation}
This theoretical connection has practical implications. When EM is implemented using gradient-based optimization in the M-step, and the E-step uses the most recent model parameters, taking a single gradient step during maximization makes the method equivalent to direct gradient-based optimization of the marginal log-likelihood.\footnote{For example, the EM-based implementation of the RegressionEM~\cite{Wang2018RegressionEM} click model in TensorFlow ranking leads to the same gradient as the binary cross-entropy loss in Eq.~\ref{eq:marginal_likelihood}, \url{https://www.tensorflow.org/ranking/api_docs/python/tfr/keras/losses/ClickEMLoss}.} More broadly, EM can be interpreted as gradient ascent with adaptive, parameter-dependent step size~\cite{Salakhutdinov2003EM}. However, we emphasize that classical implementations of both approaches differ: EM may perform full maximization in each M-step or aggregate the entire dataset in the E-step, while gradient-based methods can exploit stochasticity and adaptive optimization.

%% file: sections/4-clax.tex
\section{An Overview of CLAX}
\label{sec:clax}

We designed CLAX around three principles:
\begin{enumerate*}[label=(\roman*)]
    \item direct and numerically stable log-likelihood optimization to  replace EM;
    \item decoupling model logic and parameterization for flexible model  composition; and
    \item speed and memory-efficiency to support scale.
\end{enumerate*}
Below, we give an overview of the CLAX API and detail decisions that enable flexible parameterization and scale. We cover numerical stability separately in Section~\ref{sec:numerical-stability}.

\vspace{-0.5em}
\subsection{The CLAX model API}
All click models in CLAX share a unified interface of five methods that accept a batch of data, as demonstrated in Listing~\ref{lst:model} below:
\begin{lstlisting}[language=Python, label={lst:model}, caption={The CLAX model API.}]
batch = {
    "positions": [[1, 2, 3, ...]],
    "query_doc_ids": [[101, 205, 847, ...]], 
    "clicks": [[0., 1., 0., ...]],
    "mask": [[True, True, True, ...]],
}

loss = model.compute_loss(batch)
log_probs = model.predict_clicks(batch)
cond_log_probs = model.predict_conditional_clicks(batch)
relevance_scores = model.predict_relevance(batch)
output = model.sample(batch, rngs=nnx.Rngs(42))
\end{lstlisting}
A batch in CLAX is a Python dictionary of 2D NumPy arrays of shape (batch size, max. positions). By default, CLAX expects an array of document positions starting at 1, query-document-ids, clicks, and a binary mask. Both the variables and their names depend on the specific model parameterization and can be changed.

We require all queries within a batch to be padded to the same shape. This allows vectorizing operations across variable-length user sessions and reduces JIT recompilation when JAX encounters arrays of different lengths. A binary mask indicates which query-document pairs a model or metric should use per batch, a common pattern in Jax~\cite{Bradbury2018Jax,Jagerman2022Rax}. Each CLAX model implements five methods:
\begin{itemize}[leftmargin=*]
    \item \lstinline|compute_loss(...)| Computes the training objective, typically the negative log-likelihood of observed clicks, but custom models may implement alternative loss functions.
    \item \lstinline|predict_clicks(...)| Returns log-probabilities of a click for each document $d$ at rank $k$: $\hat{c} = P(C=1 \mid d, k)$.
    \item \lstinline|predict_conditional_clicks(...)| Returns click log-probabilities conditioned on previous clicks in the session: $\hat{c} = P(C=1 \mid d, k, c_{<k})$.
    \item \lstinline|predict_relevance(...)| Returns ranking scores for documents, typically the document attractiveness, though some models (e.g., the DBN) rank by attractiveness and satisfaction.
    \item \lstinline|sample(...)| Generates click sequences for the current batch and also returns all latent variables (examination, attractiveness, satisfaction) sampled in the process.
\end{itemize}
\vspace{-0.5em}
\subsection{Flexible parameterization}
By decoupling the structure of each click model, i.e., how variables interact with each other, from the actual parameterization, we allow flexible composition of models. CLAX supports embeddings, deep neural networks, and custom Flax models, enabling researchers and practitioners to adopt parameters to their specific use case. The following is a brief overview of parameterization options in CLAX.

\subsubsection*{Embeddings}
Click models commonly allocate separate parameters for different model components. For instance, examination parameters across positions or distinct attractiveness parameters per query-document pair. Therefore, CLAX uses embedding tables by default. Consider the UBM~\cite{Dupret2008UBM} in Listing~\ref{lst:ubm-example}. By default, the model allocates 100M embeddings for query-document attractiveness and a table of examination parameters. CLAX offers two extensions beyond traditional embedding tables that enable more accurate click predictions and scaling to larger datasets: baseline corrections and compression. Additionally, CLAX supports feature-based models as an alternative to embedding tables, which we cover afterward.

\subsubsection*{Baseline correction} Learning strictly separated parameters for attractiveness is challenging when many query-document pairs rarely occur. Therefore, CLAX optionally adds a shared baseline parameter to all embeddings in a table, so embeddings encode their offset from the global value rather than absolute values. New parameters start at the baseline and gradually adapt with more observations, which improves click prediction on long-tailed data.

\begin{lstlisting}[language=Python, label={lst:hash-embeddings}, caption={Adding hashing and baseline correction to a CCM.}]
model = ClickChainModel(
    attraction=EmbeddingParameterConfig(
        use_feature="query_doc_ids",
        parameters=100_000_000,
        add_baseline=True,
        embedding_fn=partial(
            HashEmbedding,
            compression_ratio=10
        ),
    ),
    rngs=nnx.Rngs(42),
)
\end{lstlisting}

\subsubsection*{Compression} Embedding tables can rapidly expand, exhausting memory and affecting computational efficiency. Most deep learning frameworks compute gradients for entire embedding tables, even though only the embeddings used in the current batch have non-zero gradients.\footnote{PyTorch solves this problem with sparse embedding tables and optimizers: \url{https://docs.pytorch.org/docs/stable/generated/torch.optim.SparseAdam.html}. Sparse embeddings in Jax are still under construction: \url{https://github.com/jax-ml/jax-tpu-embedding}} While GPUs mitigate this inefficiency through parallel computation, CPU-based training can slow down drastically as embedding tables grow. Thus, CLAX provides two embedding compressions:
\begin{enumerate*}[label=(\roman*)]
\item The hashing-trick \cite{Weinberger2009HashingTrick} maps multiple indices to the same embedding using hash functions, reducing table size at the cost of hash collisions.
\item The quotient-remainder trick \cite{Shi2020CompositionalEmbeddings} splits each embedding into components from two smaller tables based on the quotient and remainder of the embedding index. The final representation is a combination of both embeddings, reducing memory usage and embedding collisions.
\end{enumerate*}
Listing~\ref{lst:hash-embeddings} shows how to configure the hashing-trick for the Click Chain Model (CCM). A compression ratio of ten means that the method will hash 100M embeddings down to 10M embedding parameters. Section~\ref{sec:results:scale} evaluates both compression techniques and demonstrates training on datasets with billions of query-document pairs on a single GPU.

\subsubsection*{Feature-based models}
In many cases, allocating separate embedding parameters for models might not be optimal, e.g., because the dataset is too sparse. Instead, we might want to generalize over shared feature representations, like two-tower models that use feature representations for examination and attractiveness parameters~\cite{Guo2019PAL,Zhao2019AdditiveTowers,Yan2022TwoTowers}. CLAX supports easy configuration of any traditional click model to use features, with built-in support for linear layers, deep-neural networks, and DeepCrossV2 networks, which explicitly learn higher-order feature interactions~\cite{Wang2021DeepCrossV2}.

Listing~\ref{lst:two-towers} configures a two-tower model using a linear combination of bias features to predict examination and a DeepCrossV2 network to predict relevance from 136-dimensional feature vectors:

\begin{lstlisting}[language=Python, label={lst:two-towers}, caption={Building a two-tower model in CLAX.}]
model = PositionBasedModel(
    examination=LinearParameterConfig(
        use_feature="bias_features",
        features=8,
    ),
    attraction=DeepCrossParameterConfig(
        use_feature="query_doc_features",
        features=136,
        cross_layers=2,
        deep_layers=2,
        combination=Combination.STACKED,
    ),
    rngs=nnx.Rngs(42),
)
\end{lstlisting}

Lastly, we note that model parameters can be any Flax module, as long as the output shape of the module matches the expectations of the click model as we demonstrate in our online repository.

\vspace{-0.5em}
\subsection{Mixture models}
The power of modular design and gradient-based optimization becomes more apparent when exploring meta-modeling approaches that combine multiple click models to capture diverse user behaviors. A prominent example is the MixtureEM method proposed by \citet{Yan2022TwoTowers}, which uses the EM algorithm to learn a distribution over multiple click models, capturing different user behaviors across sessions. This approach recognizes that users may exhibit distinct browsing patterns across queries.
The original MixtureEM approach alternates between computing posterior probabilities for assigning each session to a model based on observed clicks, then training individual models with weighted losses based on these posteriors. While MixtureEM can effectively train an unbiased ranker, the posterior computation requires observed clicks, limiting click prediction on unseen rankings~\cite{Yan2022TwoTowers}.

CLAX offers a mixture model that extends the idea of \citet{Yan2022TwoTowers} and can be used like any other CLAX model. Our mixture model combines $M$ different click models, where each model $m \in M$ has a learnable prior probability $P(m)$ and produces a session-level log-loss $\mathcal{LL}_m(s)$. The loss of the mixture model is:
\begin{equation}
\mathcal{LL}_{\text{mixture}}(s) = -\log\left(\sum_{m \in M} P(m) \exp(-\mathcal{LL}_m(s) / \tau)\right),
\end{equation}
where $\tau$ is a temperature parameter controlling how much the mixture concentrates on the best-fitting models for each session. The learnable priors $P(m)$ are jointly optimized with the marginal log-likelihood of each model using gradient descent. The resulting model enables click prediction for new sessions without requiring observed clicks to compute posteriors and fully uses end-to-end gradient-based optimization. Listing~\ref{lst:mixture} shows how to learn a mixture distribution over a PBM and DBN model:
\vspace*{-0.5em}
\begin{lstlisting}[language=Python, label={lst:mixture}, caption={A mixture model with parameter sharing.}]
rngs = nnx.Rngs(42)
attraction = EmbeddingParameter(
    EmbeddingParameterConfig(
        use_feature="query_doc_ids",
        parameters=100_000_000
    ),
    rngs=rngs,
)
pbm = PositionBasedModel(
    attraction=attraction, positions=10, rngs=rngs
)
dbn = DynamicBayesianNetwork(
    attraction=attraction, positions=10, rngs=rngs
)
model = MixtureModel(models=[pbm, dbn])
\end{lstlisting}
Note that \citet{Yan2022TwoTowers} share parameters between different click models, which is not necessary in CLAX, but easy to do as we can supply the same parameter to both models, as shown in Listing~\ref{lst:mixture}. We evaluate a mixture model as part of our experiments in Section~\ref{sec:results:baidu-ultr-uva}.

\vspace{-0.5em}
\subsection{Evaluation}
More critical than training models is evaluating click models. Typically, we assess two main aspects of click models~\cite{Chuklin2015ClickModels,Grotov2015ComparativeStudy}: a model's ability to predict clicks, which is evaluated on a hold-out test set of clicks, and the model's ability to rank documents, which is assessed against relevance judgments from expert annotators. In the following, we cover the evaluation metrics implemented in CLAX.

\subsubsection*{Log-Likelihood} The most common metric for click prediction is the log-likelihood, measuring how well a model fits observed clicks:
\begin{equation}
    \operatorname{LL}(\mathcal{D}) = \frac{1}{|\mathcal{D}|} \sum_{(d, k, c) \in \mathcal{D}} \Big[ c \log \hat{c} + (1 - c) \log \left(1 - \hat{c} \right) \Big],
\end{equation}
where $\hat{c} = P(C = 1 \mid d, k, c_{<k})$ are a model's click predictions, conditioned on clicks observed before the current rank $k$. Log-likelihood values are negative, with higher values (closer to zero) indicating better model fit.

\subsubsection*{Perplexity}
Perplexity offers a more intuitive interpretation than log-likelihood. It measures how \emph{surprised} a model is by the observed data, with a lower value indicating a better model fit. Intuitively, it represents the weighted average number of choices a model is considering. Perfect predictions yield a perplexity of $1$, while random guessing for binary outcomes gives a perplexity of $2$, as the model is as uncertain as a coin flip. Perplexity is defined as:
\begin{equation}
    \operatorname{PPL}(\mathcal{D}) = 2^{- \frac{1}{|\mathcal{D}|} \sum_{(d, k, c) \in \mathcal{D}} \Big[ c \log_2 \hat{c} + (1 - c) \log_2 \left(1 - \hat{c} \right) \Big]}.
\end{equation}
Perplexity metrics differ based on how the click prediction $\hat{c}$ is calculated:
\begin{enumerate*}[label=(\roman*)]
    \item Conditional perplexity uses $\hat{c} = P(C=1 \mid d, k, c_{<k})$, where predictions can incorporate clicks observed at previous ranks.
    \item Unconditional perplexity uses $\hat{c} = P(C=1 \mid d, k)$, without considering clicks from the current session.
\end{enumerate*}
This distinction is important, since some click models adapt their predictions based on clicks in the current sessions (such as ~\cite{Dupret2008UBM}). Conditional perplexity measures how well a model fits an observed dataset, while unconditional perplexity reflects how accurately a model can predict clicks on a completely unseen list of documents. Note that conditional perplexity assumes top-down browsing behavior; when this assumption is violated, unconditional perplexity is preferable.

CLAX implements all three standard click prediction metrics with support for global and rank-based averaging. Similar to the model API, all metrics handle batched inputs with a binary mask indicating which input document are not padding. Our API follows the NNX metrics API\footnote{\url{https://flax.readthedocs.io/en/latest/api_reference/flax.nnx/training/metrics.html}} and supports input routing, meaning all metrics can be updated at once with each metric automatically extracting the arguments it requires, as shown in Listing~\ref{lst:metrics} below:
\begin{lstlisting}[language=Python, label={lst:metrics}, caption={Computing click metrics.}]
metrics = MultiMetric({
    "ll": LogLikelihood(),
    "ppl": Perplexity(),
    "cond_ppl": ConditionalPerplexity(),
})

metrics.update(
    log_probs=log_probs,
    conditional_log_probs=cond_log_probs,
    clicks=clicks,
    where=mask,
)

results = metrics.compute()
rank_results = metrics.compute_per_rank()
\end{lstlisting}

\subsubsection*{Ranking metrics}
A second aspect commonly evaluated of click models is their ranking performance, which (in web search) is typically assessed against expert-annotated relevance labels using metrics such as discounted cumulative gain (DCG), mean reciprocal rank (MRR), or average precision (AP). Instead of reimplementing these metrics, CLAX supports integrating ranking metrics from Rax, the most prevalent JAX-based library for learning-to-rank~\cite{Jagerman2022Rax}:

\vspace{-0.5em}
\begin{lstlisting}[language=Python, caption=Support for Rax-based ranking metrics.]
metrics = MultiMetric({
    "dcg@10": RaxMetric(rax.dcg_metric, top_n=10),
    "mrr@10": RaxMetric(rax.mrr_metric, top_n=10),
})

metrics.update(scores=scores, labels=labels, where=mask)
results = metrics.compute()
\end{lstlisting}
Following this overview of CLAX, we introduce the basic techniques used to achieve numerical stability in this work.

%% file: sections/5-numerical-stability.tex
\vspace{-0.5em}
\section{Numerical Stability}
\label{sec:numerical-stability}

Optimizing complex likelihood expressions using gradient-based optimization requires attention to numerical stability. The marginal likelihoods of many common click models contain products of small probabilities, which can lead to numerical underflow in finite-precision computer arithmetic~\cite{Goldberg1991Floats,Kahan1996IEEE}. Below, we cover the techniques CLAX uses to stabilize complex likelihood expressions by performing all probability computations in log-space.

\subsubsection*{Multiplication} By moving to log-probabilities, products of probabilities simplify to sums (and division to subtraction):
\begin{equation}
    \log \left( \prod_{i = 1}^{n} p_i \right) = \sum_{i = 1}^{n} \log p_i,
\end{equation}
which essentially eliminates the concern of numerical underflow when multiplying small probabilities.

\subsubsection*{Addition} While multiplication becomes more stable (and faster) in log-space, the addition of probabilities becomes more complicated as it requires first exponentiating log probabilities. This reintroduces the instabilities we seek to avoid, as exponentiating large positive inputs lead to overflow and exponentiating large negative inputs lead to underflow. The standard solution is to avoid large inputs to the $\exp(\cdot)$ operation via the log-sum-exp trick~\cite{Blanchard2019LogSumExp}:
\begin{equation}
\texttt{log\_sum\_exp}(a) = a_{\text{max}} + \log \left( \sum_{i=1}^{n} \exp(a_i - a_{\text{max}}) \right),
\end{equation}
where $a = (a_1, \dots, a_n)$ is a vector of log values and $a_{\text{max}} = \max_i(a_i)$ is the maximum input value. The trick is prevalent in probabilistic modeling, and we also use it to transform the output logits of neural networks $x \in \mathbb{R}$ to log-probabilities by implementing numerically stable versions of the log-sigmoid functions:
\begin{equation}
    \begin{split}
        \log(\sigma(x)) &= - \texttt{log\_sum\_exp}([0, -x]) \text{, or }\\
        \log(1 - \sigma(x)) &= - \texttt{log\_sum\_exp}([0, x]).\\
    \end{split}
\end{equation}

\subsubsection*{Complements and cancellation} Sometimes we need to compute the log of a complement $\log(1 - p)$, e.g., in the binary-cross entropy loss or when computing log-posteriors in the DBN~\cite{Chapelle2009DBN}. Performing this step directly from log-probability $\log p$ requires computing: $\log(1 - \exp(\log p))$. This expression is numerically unstable in two ways:
\begin{enumerate*}[label=(\roman*)]
\item underflow: when $p$ is very small, $\log p$ is very negative, causing $\exp(\log p)$ to underflow to zero; and
\item catastrophic cancellation: when $p \approx 1$, we have 
$\exp(\log p) \approx 1$, making $1 - \exp(\log p) \approx 0$, since subtracting nearly equal floating point numbers leads to a loss of precision~\cite{Goldberg1991Floats}.
\end{enumerate*}
Therefore, we compute $\texttt{log1mexp}(x)$ as proposed in \citep{Maechler2012Log1mExp} and adopted by major frameworks such as TensorFlow\footnote{\url{https://www.tensorflow.org/probability/api_docs/python/tfp/math/log1mexp}} and JAX.\footnote{\url{https://docs.jax.dev/en/latest/_autosummary/jax.nn.log1mexp.html}} \citet{Maechler2012Log1mExp} proposes a piecewise approximation that switches between two stable expressions that are precise in different input ranges.\footnote{We direct interested readers for the theoretical motivation behind the switching point $\log(2)$ to \cite[Section 2]{Maechler2012Log1mExp}.} For a log-probability $a \in \mathbb{R}, a \leq 0$:
\begin{equation}
    \texttt{log1mexp}(a) =
    \begin{cases}
        \log(-\text{expm1}(a)) & \text{if } a > -\log(2) \\
        \text{log1p}(- \exp(a)) & \text{if } a \le -\log(2).
    \end{cases}
\end{equation}
The implementation relies on the standard functions $\text{log1p}(x)$, which accurately computes $\log(1 + x)$, and $\text{expm1}(x)$, which accurately computes $\exp(x) - 1$, to avoid catastrophic cancellation.

To summarize, CLAX performs all probability computations in log space for increased numerical stability, avoiding underflow and overflow as well as catastrophic cancellation. We list all implemented log-likelihoods in Appendix~\ref{appendix:models} and their corresponding implementation can be found in our code repository.\footnote{\url{https://github.com/philipphager/clax}}

%% file: sections/6-experiments.tex
\begin{figure*}[ht]
    \includegraphics[width=1\textwidth]{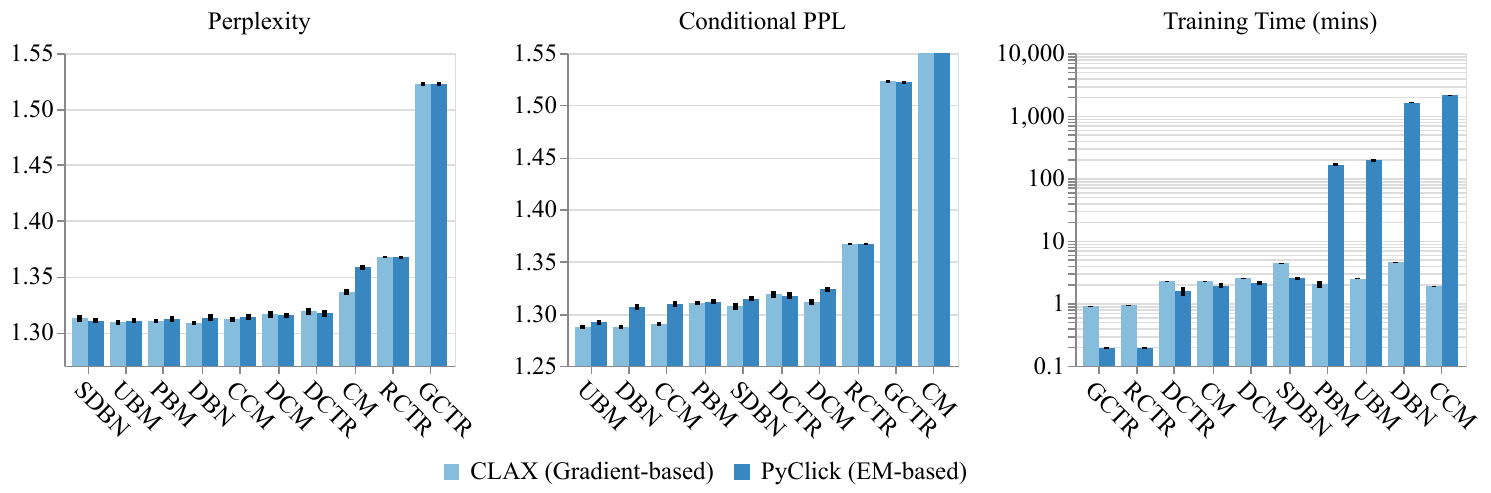}
    \caption{CLAX matches or exceeds the click predictions of PyClick over three folds of 10M training sessions on WSCD-2012.}
    \label{fig:1-yandex-baseline}
    \vspace{-1em}
\end{figure*}

\begin{figure}[ht]
    \includegraphics[width=1\columnwidth]{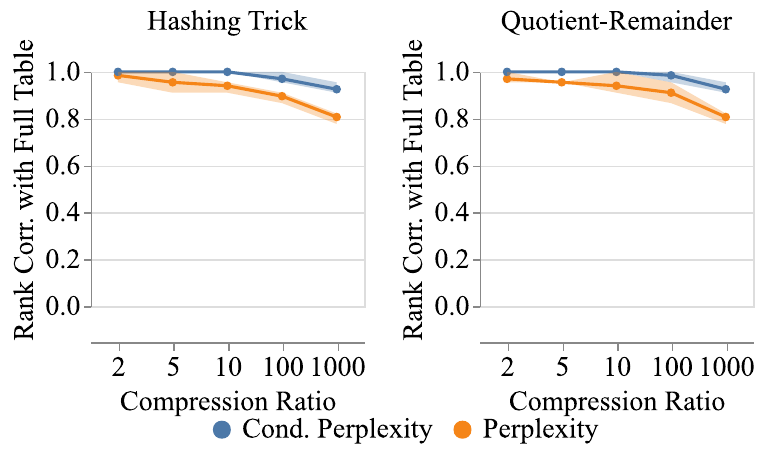}
    \caption{Kendall's $\tau$ between ranking models trained with and without embedding compression on WSCD-2012.}
    \label{fig:2-yandex-compression}
    \vspace*{-2em}
\end{figure}

\vspace{-0.5em}
\section{Experimental Setup}
\label{sec:experiments}

We conduct experiments in three settings to evaluate CLAX. First, we compare CLAX models to EM-based counterparts from PyClick. Second, we scale CLAX to large datasets and investigate the effects of embedding compression. Third, we evaluate CLAX models as unbiased ranking models. Next, we introduce the basic setup shared across all our experiments.

\subsubsection*{Datasets} We use two real-world datasets of user interactions with search engines: The WSCD-2012 dataset by Yandex is a foundational benchmark in click modeling~\cite{Serdyukov2012WSCD,Chuklin2015ClickModels}. It contains 146,278,823 user sessions and 346,711,929 unique query-document pairs. The dataset provides query and document identifiers without additional document features, allowing only for a direct comparison of embedding-based click models. We generate a unique identifier for each query-document combination as the only preprocessing step.

The Baidu-ULTR dataset is the largest real-world dataset for unbiased learning-to-rank, comprising over 1.2 billion user sessions and a test set of 397,572 annotated query-document pairs~\cite{Zou2022Baidu}. This scale allows us to verify CLAX's scalability and ranking performance. We employ hashing to generate query-document IDs from query IDs and document URLs, yielding 2,147,483,647 unique identifiers. We are the first to train on the entire Baidu-ULTR dataset, rather than just a subset~\cite{Zou2022Baidu,Hager2024ULTR,WSDMCup2023ULTR,Li2023-TENCENT-PRETRAIN-1,Chen2023-TENCENT-ULTR-1}. For ranking performance comparison, we use the subset of Baidu-ULTR created by \citet{Hager2024ULTR} with pre-computed 768-dimensional MonoBERT features for 2,372,947 sessions.\footnote{\url{https://huggingface.co/datasets/philipphager/baidu-ultr_baidu-mlm-ctr}} We publish all pre-processed datasets and highly efficient custom dataloaders using Apache Parquet under \url{https://huggingface.co/datasets/philipphager/clax-datasets}.

\subsubsection*{Implementation} All CLAX experiments use the default trainer with the AdamW optimizer~\cite{Loshchilov2019AdamW} (learning rate 0.003, weight decay 0.0001) over 100 epochs, stopping early after the first epoch without improvement of the validation loss. Beyond our preliminary experiments, hyperparameters should be tuned per model and dataset. All experiments run over three dataset splits and random seeds, we plot bootstrapped 95\% confidence intervals. CLAX experiments use a single NVIDIA RTX A6000 GPU (48GB RAM) and PyClick experiments use an Intel Xeon Gold 5118 CPU (2.30GHz). As recommended, we run PyClick on the PyPy interpreter with JIT-compilation. To ensure a fair comparison with CLAX, we adjust PyClick's parameter initialization from $\frac{1}{2}$ to $\frac{1}{9}$ to better reflect the mean CTR on WSCD-2012, improving click prediction on long-tailed items. We publish all experimental configurations including detailed data splits and baselines in our repository: \url{https://github.com/philipphager/clax}

%% file: sections/7-results.tex
\begin{figure*}[th]
    \includegraphics[width=1\textwidth]{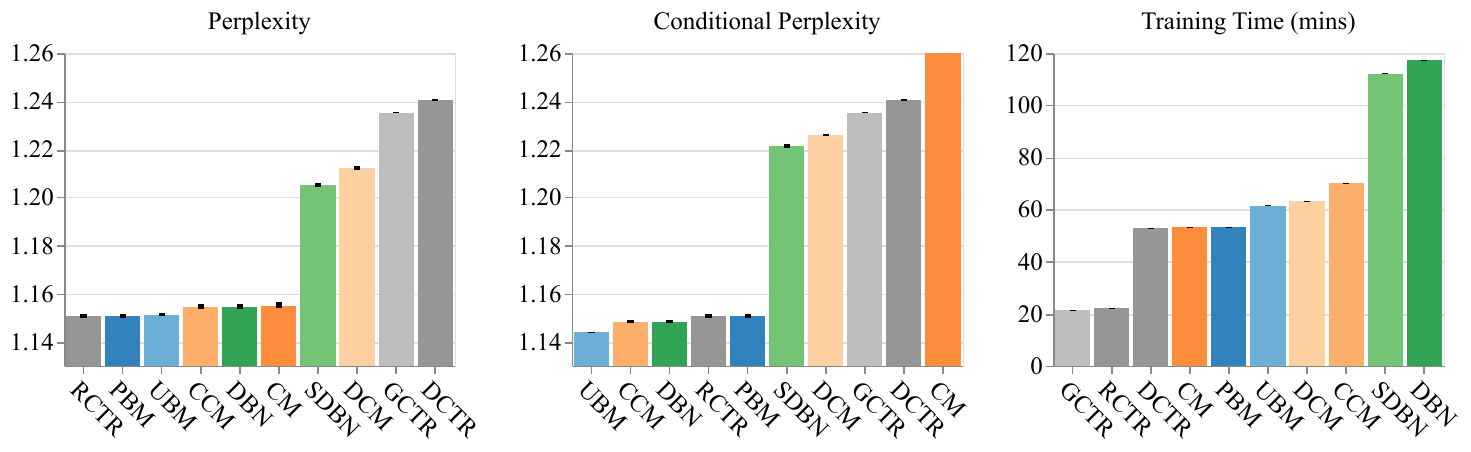}
    \caption{Embedding-based CLAX models on the Baidu-ULTR dataset (three folds of 800M / 200M / 200M sessions for training,  validation, and testing)~\cite{Zou2022Baidu}. All models complete training under 2 hours using the hashing-trick with 10x compression.}
    \label{fig:3-baidu-ultr-embeddings}
    \vspace{-0.5em}
\end{figure*}
\begin{figure*}[th]
    \includegraphics[width=1\textwidth]{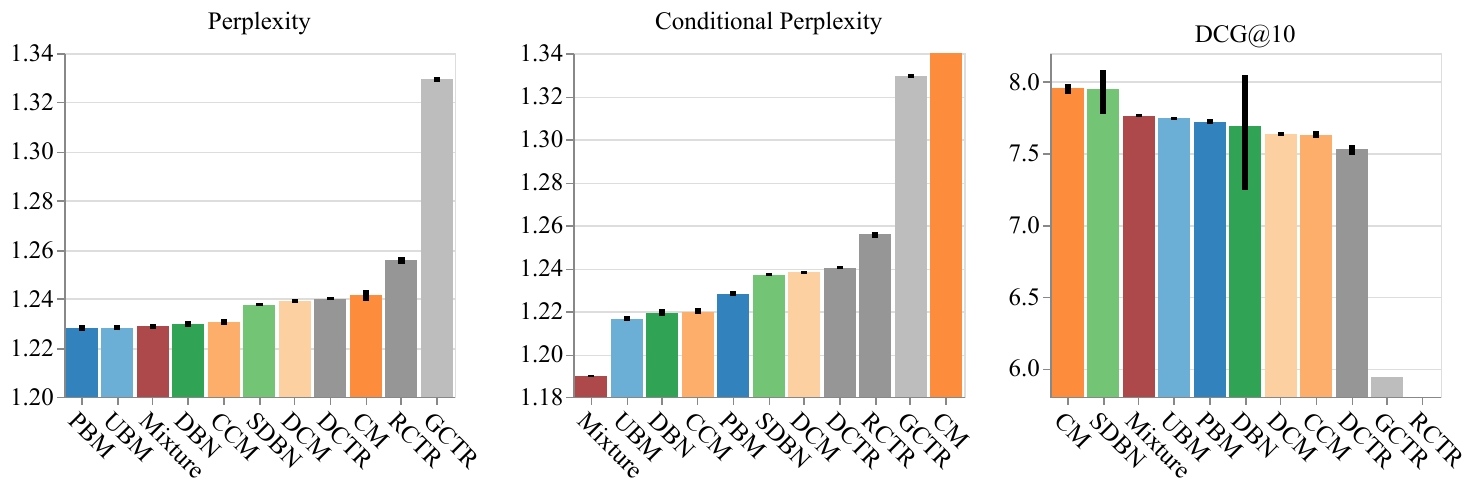}
    \caption{CLAX models generalizing over BERT features on the Baidu-ULTR-UVA dataset~\cite{Hager2024ULTR} using a deep-cross network achieve strong ranking performance and a different model fit compared to embedding-based models.}
    \label{fig:4-baidu-ultr-features}
    \vspace{-1em}
\end{figure*}

\vspace*{-0.5em}
\section{Results}
\label{sec:results}

\subsubsection*{Comparing EM and gradient-based optimization}
\label{sec:results:yandex}

We validate CLAX against PyClick, the prevailing click modeling library, to ensure our implementation produces equivalent click predictions. Due to the scalability limitations of PyClick, we train on three folds of 10M user sessions from the WSCD-2012 dataset, using 5M sessions each for validation and testing. Figure~\ref{fig:1-yandex-baseline} compares the click prediction performance and training times across both libraries. First, we note that the (unconditional) perplexity matches closely between the two libraries. For conditional perplexity, simple models, such as the PBM and CTR-based models, achieve identical performance. Surprisingly, CLAX sometimes achieves better conditional perplexity despite both libraries optimizing the same objectives. After further investigation, we attribute this improvement to CLAX's enhanced numerical stability, as we find improved click predictions at lower ranks. Regarding training time, PyClick excels for MLE-based models, which just require counting. Training the EM-based models on 10M sessions, however, requires from 172 minutes (PBM) to 36 hours (CCM).\footnote{Note that the DCM in PyClick is actually a simplified DCM (SDCM) to use faster MLE while CLAX implements the original latent-variable version~\cite{Guo2009DCM}.} In contrast, all CLAX models complete training in under 5 minutes. To summarize, CLAX matches PyClick's unconditional click prediction performance while matching or exceeding conditional prediction accuracy, potentially leading to different conclusions about optimal model selection for specific datasets.

\vspace{-0.5em}
\subsubsection*{Scaling up CLAX}
\label{sec:results:scale}
Next, we evaluate CLAX on large datasets. To begin, we evaluate the hashing trick and quotient-remainder embedding compressions. To assess how compression might change our conclusions about model fit, we compare the obtained model ranking when training with compression versus training on the full embedding table. We train on WSCD-2012 using three splits of 80M training sessions and evaluate five compression ratios, reducing the full embedding table with 346M entries by factors of 2x to 1,000x.

Figure~\ref{fig:2-yandex-compression} shows the resulting Kendall's tau rank correlation when sorting models by their click prediction performance, trained with compression, against the ranking obtained when training without compression. We observe that both compression methods behave similarly, maintaining remarkably high correlation up to very high compression ratios of 10-100x. Unconditional perplexity is more susceptible to compression. However, as Figure~\ref{fig:1-yandex-baseline} reveals, many models perform similarly on this dataset, so small changes can alter rankings while preserving the overall conclusion that many models are nearly equivalent. Note that compression reduces overall click prediction performance (higher perplexity). As some CTR baselines, such as the RCTR and GCTR, do not use compression, comparing compressed and uncompressed models can lead to wrong conclusions at high compression rates. We observe that, beyond reducing the memory footprint, compression also decreases the average training time from 14 minutes for uncompressed models to around four minutes for 10x compression and higher.
Secondly, we evaluate scale by training CLAX models using the hashing-trick with 10x compression on three folds of the full Baidu-ULTR dataset containing over 1B user sessions. Fig.~\ref{fig:3-baidu-ultr-embeddings} shows the resulting models, all completing training under 2 hours.

\vspace{-0.5em}
\subsubsection*{Generalizing over features}
\label{sec:results:baidu-ultr-uva}

Lastly, we parameterize CLAX's attractiveness and satisfaction parameters with a deep-cross network to investigate click prediction and ranking performance when generalizing over query-document features. Figure~\ref{fig:4-baidu-ultr-features} shows the results on the Baidu-ULTR-UVA subset~\cite{Hager2024ULTR}. While click prediction performance remains similar to that of embedding-based training, the performance gap between individual models narrows considerably when generalizing over features, leading to different conclusions about model relationships. For ranking performance, we focus on the DCTR model (corresponding to a naive model without bias correction in unbiased learning-to-rank) and PBM (corresponding to a two-tower model). Ranking performance on Baidu-ULTR does not directly correlate with click prediction performance, a known problem on this dataset~\cite{Hager2024ULTR}. Nevertheless, cascade-based models achieve strong ranking performance, comparable to listwise LTR loss functions trained in previous work~\cite[Figure 3]{Hager2024ULTR}. Our results suggest that complex click models can be effective ranking models.

Finally, we evaluate the effectiveness of our mixture model, which combines a PBM, DCTR, and GCTR model, in Figure~\ref{fig:4-baidu-ultr-features}, following the setup of \citet{Yan2022TwoTowers} but excluding the RCTR as it cannot be applied to the Baidu-ULTR test. The mixture model achieves better model fit and ranking performance than each individual model.

%% file: sections/8-conclusion.tex
\vspace{-0.5em}
\section{Conclusion}
\label{sec:conclusion}

We have introduced CLAX, the first JAX-based click modeling library enabling end-to-end gradient-based optimization for PGM-based click models. CLAX demonstrates that gradient-based optimization can replace EM for training click models while achieving comparable performance. The framework provides orders of magnitude speedup over established implementations, training and evaluating on over 1B user sessions in $\approx 2$ hours on a single GPU.

CLAX's modular design decouples model logic from parameterization, supporting embeddings, deep networks, and custom modules. Through embedding compression techniques, the framework scales to billions of query-document pairs. Our experiments show that neural parameterizations of complex PGMs and mixture models can surpass widely-used two-tower models in ranking tasks.

The CLAX framework serves both industry practitioners seeking to understand user behavior at scale and researchers developing new click models. All code and datasets are open-source, enabling reproducible research and practical adoption.

Our implementation currently lacks support for sparse embeddings, which can negatively impact performance on CPUs. In the future, we will support sparse embeddings, neural click models beyond classical PGMs, and distributed training across GPUs.

%% file: sections/9-acknowledgements.tex
\vspace*{-1mm}
\begin{acks}
We thank David Rohde, Oscar Ramirez, Mathijs de Jong, and Ahmed Khaili for their invaluable feedback. This research was supported by the Mercury Machine Learning Lab created by TU Delft, the University of Amsterdam, and Booking.com.
Maarten de Rijke was supported by the Dutch Research Council (NWO), project nrs 024.004.022, NWA.1389.20.\-183, and KICH3.LTP.20.006, and the European Union's Horizon Europe program under grant agreement No 101070212. All content represents the opinion of the authors, which their employers or sponsors do not necessarily endorse.
\end{acks}

%% file: sections/10-appendix.tex
\appendix

\section{Models}
\label{appendix:models}

We introduce the ten click models implemented in CLAX following the standard work of \citet[Table 3.1]{Chuklin2015ClickModels} and their click log-probabilities. If we do not explicitly state a conditional click probability, unconditional and conditional click probabilities are equal for the specific model.

\subsection{Global CTR model}
The global CTR (GCTR) model, sometimes called the random click model, predicts a single average CTR across all documents. It serves as a simple baseline that any click model should surpass:
\begin{equation}
    \log P(C = 1 \mid d, k) = \log \rho.
\end{equation}

\subsection{Rank-based CTR model (RCTR)}
The rank-based CTR (RCTR) model assumes that click probability depends only on the rank $k$ of a document and not its content. The model predicts the average CTR for all documents displayed at the same rank, treating them as equally attractive:
\begin{equation}
    \log P(C = 1 \mid d, k) = \log \theta_k.
\end{equation}

\subsection{Document-based CTR model (DCTR)}
The document-based CTR (DCTR) model assumes that clicks depend solely on a document and not on its position in the ranking:
\begin{equation}
    \log P(C = 1 \mid d, k) = \log \gamma_d.
\end{equation}

\subsection{Position-based model (PBM)}
The PBM assumes that clicks occurs only if a user first examines the result at rank $k$ (with probability $\theta_k$), and if the displayed document is attractive ($\gamma_d$):
\begin{equation}
    \log P(C = 1 \mid d, k) = \log \theta_k + \log \gamma_{d}.
\end{equation}

\subsection{Cascade model (CM)}
The cascade model (CM) assumes that users scan results from top to bottom, click on the first attractive document they find, and then immediately stop their search. The probability of a click at rank $k$ depends on the displayed document $d$ being attractive ($\gamma_d$) and all preceding documents being unattractive:
\begin{equation}
    \log P(C=1 \mid d, k) = \log \gamma_d + \sum_{i=1}^{k-1} \log(1 - \gamma_{d_i}).
\end{equation}
Note that the cascade model can only explain a single click per list. All other documents after the first click, by definition, have a click probability of $0$. To avoid a log-likelihood of $-\infty$ in our conditional click predictions, we follow the common practice to assign a very small default click probability to all documents following a click~\cite{Chuklin2015ClickModels}:
\begin{equation}
    \log P(C=1 \mid d, k, c_{<k}) =
    \begin{cases}
        \log \gamma_d & \text{if } \sum_{i=1}^{k-1} c_i = 0 \\
        \text{min\_log\_prob} & \text{otherwise.}
    \end{cases}
\end{equation}

\subsection{User browsing model (UBM)}
The user browsing model (UBM) extends the PBM by assuming that the probability of examination at position $k$ depends also on the position of the last clicked document $k'$. This is most easily demonstrated in the conditional log-probability of click:
\begin{equation}
    \log P(C=1 \mid d, k, c_{<k}) = \log \theta_{k, k'} + \log \gamma_{d},
\end{equation}
where $k$ is the position of the current document and $k'$ the position of the previously last clicked document. While conditional click probabilities are very simple, predicting clicks on a new list of documents is harder under the UBM, since it requires marginalizing over all possible last click positions $i < k$ before our current position:
\begin{equation}
\begin{split}
    &\log P(C = 1 \mid d, k) =\\
    &\log \left( \sum_{i=0}^{k - 1} P(C=1 \mid d_i, i) \cdot \left(\prod_{j=i+1}^{k - 1} (1 - \theta_{j,i}\gamma_{d_j})\right)  \theta_{k,i}\gamma_{d} \right).
\end{split}
\end{equation}
Each term in the sum represents a path to the current document: the probability of clicking at a previous rank $i$, then not clicking on anything until rank $k$, and finally examining and clicking the document at rank $k$ given $i$ was the last clicked position.

\subsection{Dependent click model (DCM)}
The dependent click model (DCM) is an extension of the cascade model to explain multiple clicks in a single ranking. The DCM assumes that users examine a list from top to bottom, click on relevant items, and after clicking have a rank-dependent probability $\lambda_k$ to continue browsing:
\begin{equation}
    \begin{split}
    \log P(C=1 \mid d, k) &= \log(\epsilon_{k}) + \log(\gamma_{d})\\
    \log(\epsilon_{k+1}) &= \log(\epsilon_k) + \log(\gamma_{d_k} \lambda_k + (1 - \gamma_{d_k})).\\
    \end{split}
\end{equation}
When conditioning on observed clicks, the examination probability changes based on the actions in the current session:
\begin{equation}
    \begin{split}
    \log P(C=1 \mid d, k, c_{<k}) &= \log(\epsilon_{k}) + \log(\gamma_{d})\\
    \log(\epsilon_{k+1}) &= \log\left(c_k \lambda_k + (1 - c_k) \frac{(1 - \gamma_{d_k}) \epsilon_k}{1 - \gamma_{d_k} \epsilon_k}\right).\\
    \end{split}
\end{equation}
If a user clicks on a document, they continue to the next rank with probability $\lambda_k$ and if they do not click, we calculate the posterior probability of examining the next rank given that we observed no click using Bayes' rule.

\subsection{Click chain model (CCM)}
The click chain model (CCM) is an extension of the DCM, assuming a total of three continuation scenarios that do not only explain continuation after clicking a document but also allow users to abandon a session without any clicks. First, $\tau_1$ is the probability of a user continuing to the next document after not clicking on the current document. Second, if the user clicks on the current document but is not satisfied, $\tau_2$ is the probability of the user continuing to the next position. And lastly, $\tau_3$ is the probability that a user clicks on the current item, finds it satisfying, but still wants to continue to the next document:
\begin{equation}
    \begin{split}
    \log P(C=1 \mid d, k) &= \log(\gamma_d) + \log(\epsilon_k) \\
    \log(\epsilon_{k+1}) &= \log(\epsilon_k) \\
    &\quad + \log\left( \gamma_{d_k}((1-\gamma_{d_k})\tau_2 + \gamma_{d_k}\tau_3) \right. \\
    &\quad \left. + (1-\gamma_{d_k})\tau_1 \right).
    \end{split}
\end{equation}
When conditioning on the observed clicks, the update rule for the examination probability changes based on the user's action at the current rank. If a click occurred, we compute continuation based on satisfaction (equal to attractiveness $\gamma_d$) and the continuation probabilities $\tau_2$ and $\tau_3$. If no click was observed, we compute the posterior log probability of continuing to the next rank:
\begin{equation}
\begin{split}
\log P(C=1 \mid d, k, c_{<k}) &= \log(\gamma_d) + \log(\epsilon_k) \\
\log(\epsilon_{k+1}) &= c_k \left[ \log\left(\gamma_{d_k}\tau_3 + (1-\gamma_{d_k})\tau_2 \right) \right] \\
&\quad + (1-c_k) \left[ \log(1-\gamma_{d_k}) + \log(\epsilon_k) \right. \\
&\quad \left. + \log(\tau_1) - \log(1 - \gamma_{d_k}\epsilon_k) \right].
\end{split}
\end{equation}

\subsection{Dynamic Bayesian network}
The dynamic Bayesian network (DBN) model separates the concepts of a document being attractive ($\gamma_d$) and being satisfying ($\sigma_d$). A user stops their search only if they click on an attractive document and are satisfied by it. If they do not click or are not satisfied by the clicked document, they continue browsing with a global continuation probability $\lambda$:
\begin{equation}
    \begin{split}
    \log P(C=1 \mid d, k) &= \log(\gamma_d) + \log(\epsilon_k) \\
    \log(\epsilon_{k+1}) &= \log(\epsilon_k) + \log(\lambda) + \log(1 - \gamma_{d_k}\sigma_{d_k}).\\
    \end{split}
\end{equation}
The conditional click probability again takes the user's actions in the current session into account. If a click was observed, we compute the probability of continuation based on satisfaction. If no click was observed, we compute the posterior probability of continuing to the next item:
\begin{equation}
    \begin{split}
    \log P(C=1 \mid d, k, c_{<k}) &= \log(\gamma_d) + \log(\epsilon_k) \\
    \log(\epsilon_{k+1}) &= \log(\lambda) + c_k \left[ \log(1 - \sigma_{d_k}) \right] \\
    &\quad + (1-c_k)\left[ \log(1-\gamma_{d_k}) + \log(\epsilon_k) \right. \\
    &\quad \left. - \log(1 - \gamma_{d_k}\epsilon_k) \right].
    \end{split}
\end{equation}
